\RequirePackage{fixltx2e}
\documentclass[aps,pra,10pt,showpacs,twocolumn,superscriptaddress,notitlepage]{revtex4}

\usepackage[utf8]{inputenc}
\usepackage[brazilian,british]{babel}
\usepackage[T1]{fontenc}
\usepackage{amsmath,amssymb,amsthm}
\usepackage[sc,osf]{mathpazo}
\usepackage[pdftex]{graphicx}
\usepackage[update,prepend]{epstopdf}
\usepackage{verbatim}
\usepackage{xcolor}
\usepackage{bm}
\usepackage[babel=true]{microtype}
\usepackage{pifont}
\usepackage{booktabs}

\usepackage{dcolumn}

\usepackage{float}

\usepackage{ulem}

\newcommand{\bra}[1]{\left\langle #1 \right|}
\newcommand{\ket}[1]{\left| #1 \right\rangle}
\newcommand{\braket}[2]{\left\langle #1 \middle| #2 \right\rangle}
\newcommand{\ketbra}[2]{\left|#1\middle\rangle\middle\langle#2\right|}

\newcommand{\abs}[1]{\left|#1\right|}

\newcommand{\comm}[2]{\left[#1,#2\right]}

\newcommand{\ad}[1]{a^{\dagger}_{#1}}

\begin{document}
\title{Dynamical matrix for arbitrary quadratic fermionic bath Hamiltonians and non-Markovian dynamics of one and two qubits in an Ising model environment} 
\author{Fernando Iemini}
\email{fernandoiemini@gmail.com}
\affiliation{Abdus Salam ICTP, Strada Costiera 11, I-34151 Trieste, Italy}
\author{Leonardo da Silva Souza}
\affiliation{Departamento de F\'{\i}sica - ICEx - Universidade Federal de Minas Gerais,
Av. Pres. Ant\^onio Carlos 6627 - Belo Horizonte - MG - Brazil - 31270-901.}
\author{Tiago Debarba} \affiliation{Universidade Tecnol\'ogica Federal do Paran\'a (UTFPR), Campus Corn\'elio Proc\'opio, 
Avenida Alberto Carazzai 1640, Corn\'elio Proc\'opio, Paran\'a  86300-000, Brazil}
\author{Andr\'e T. Ces\'{a}rio}
\author{Thiago O. Maciel}
\author{Reinaldo O. Vianna}
\affiliation{Departamento de F\'{\i}sica - ICEx - Universidade Federal de Minas Gerais,
Av. Pres. Ant\^onio Carlos 6627 - Belo Horizonte - MG - Brazil - 31270-901.}

\date{\today}

\begin{abstract} 
 We obtain the  analytical expression for the {Kraus decomposition of the quantum map of an environment modeled by an arbitrary quadratic fermionic Hamiltonian  acting on one or two qubits, and  derive simple functions to  check the non-positivity of the intermediate map. These functions correspond to  two different sufficient criteria
for non-Markovianity. In the particular case of an environment represented by the Ising  Hamiltonian,  we discuss the two sources of non-Markovianity in the model, one due to the finite size of the lattice, and another due to the kind of interactions.}
\end{abstract}

\pacs{   05.50.+q, 03.65.Ta, 03.65.Yz, 05.70.Jk    }
\maketitle

\section{Introduction}

The need to fight decoherence, to guarantee the proper working of the quantum enhanced technologies of information and
computation \cite{Nielsen}, has renovated the 
motivation for the in-depth study of  system-environment interaction dynamics.
In particular, the Markovian or non-Markovian nature of 
the dynamics is of great interest \cite{maniscalco}. 
Several witnesses and quantifiers have been proposed 
in order to characterize the non-Markovianity of a quantum processes \cite{RHP, BLPV, Haikka11}.
For instance, the information flow between system 
and environment, quantified by the distinguishability of any two quantum states 
\cite{BLP, Chrus14, Haikka12}, or by the 
Fisher information \cite{Lu10}, or mutual information \cite{Luo12}.
Another interesting quantifier is the entanglement based measure of non-Markovianity \cite{Rivas10}. It is 
related to the classical information flow between system and environment \cite{fanch}. 
The physical interpretation of these quantifiers, however,  
 remains  an open question. The behavior of the quantifiers
 depend on the kind of interactions and  size of the system, as is discussed in  \cite{fanch2}.

In this work we wish to obtain explicitly the Choi representation of the quantum map of an arbitrary quadratic fermionic Hamiltonian acting on qubits, and perform a comparative exploration of its dynamics from the point of view of
(non-)divisibility \cite{RR11, Devi12}.  After obtaining the analytical expression of the dynamical matrix, we specialize to the case of an environment represented by  the quantum one-dimensional Ising model acting on one central qubit, which in the case of finite size lattices can be solved analytically by means of the well known Jordan-Wigner and Bogoliubov transformations \cite{Sachdev, Franchini}. The availability of an analytical solution for this representative critical  model  is the reason why
this system is recurrently investigated in many instances. The study we perform here is complementary to previous investigations and, besides its pedagogical purpose, 
reveals functional dependencies among different indicators of non-Markovianity, and also stresses that there are two sources of non-divisibility in the dynamics, one intrinsic
to the kind of interactions, and  another due to the finite size of the lattice.
The divisibility criterion  consists in checking if an intermediate quantum map is not Complete Positive (CP) for some time instant, which amounts to checking the non-positivity
of the corresponding dynamical matrix \cite{Zyc06}.  We will show that the non-positivity of the dynamical matrix, measured by its eigenvalues,  in this case is a simple function of the Loschmidt echo \cite{echo}, a quantity that indicates decoherence induced by perturbations. We will also investigate the action of a trivial extension of the map on the decay of entanglement of the system coupled to an ancilla. We shall see that the intermediate map is not contractive,  and entanglement is again a function of the Loschmidt echo which is not
monotonically decreasing, signaling non-Markovianity and information flux from the environment to the system  \cite{fanch}. Finally we wish to know if the number of particles in the 
system has some influence on the dynamics of the environment. Thus we derive the map acting on a system composed of two qubits, concluding that the results do not have
any change.

The paper is organized as follows. We briefly revise the formalism of dynamical maps and the divisibility criterion in Sec. II.  Our first result appears in Sec. III, where we 
present the {exact Kraus decomposition } for {general} quadratic  fermionic Hamiltonians, and introduce a measure of non-Markovianity. In Sec. IV, we introduce the model we shall investigate
numerically, and relate it to the formalism of Sec. III.  In Sec. V, we obtain the map for a system of two-qubits, showing that the results related to non-Markovianity do not change
in relation to the one-qubit case.
Our results for the dynamics of a qubit interacting
 with an environment governed by the Ising model are presented in 
Secs. VI and VII, where we investigate the non-Markovianity both at and outside of the critical point.
  In Sec.VI we investigate the non-Markovianity using the most negative eigenvalue of the intermediate map as a quantifier,  while in Sec. VII we use the increase of
  entanglement under local CP maps as a quantifier.  Our final remarks are in Sec. VIII.

\section{Quantum Dynamical Maps and the Divisibility Criterion }
\label{quantum.dynamical}

The evolution of an open quantum system  ($\rho^\prime=\Phi(\rho)$) can be written in the well known operator sum representation as \cite{Nielsen,Zyc06}:
\begin{equation}
\label{krausrepresentation}
\rho^\prime = \sum_\mu K_\mu \rho K_{\mu}^{\dagger},\qquad \sum_{\mu}K_{\mu}^{\dagger}K_{\mu}=\mathbb{I},
\end{equation}
where the $K_\mu$ are the Kraus operators related to the quantum map $\Phi$, and $\mathbb{I}$  is the identity in the 
Hilbert space of the system.
Using the { \it vec} operation, defined by \cite{Zyc06}:
\begin{equation}
vec(\ketbra{x}{y})= \ket{x}\otimes\ket{y},
\end{equation}
and the corresponding inverse operation,
\begin{equation}
vec^{-1}( \ket{x}\otimes\ket{y}) =\ketbra{x}{y},
\end{equation}
 the following matrix product  ($ABC$) can be cast as:
\begin{equation}
ABC=vec^{-1}[(A\otimes C^{T}) vec(B)],
\end{equation}
Therefore, a product of three matrices can be thought of as a super-operator (or map) $A\otimes C^T$ acting on the linear operator $B$.
Now  Eq.(\ref{krausrepresentation}) can be conveniently rewritten as \cite{Zyc06}:
\begin{equation}
\ket{\rho'} = \Phi \ket{\rho},\,\,\,\Phi = \sum_{\mu}K_\mu\otimes K_{\mu}^{*}\, ,
\end{equation}
where $\ket{\rho}\equiv vec(\rho)$.

Consider the evolution of the system from an initial time $t_0$ to a final time $t_f$,
\begin{equation}
\ket{\rho(t_f)} = \Phi(t_f,t_0) \ket{\rho(t_0)}.
\end{equation} 
Suppose this evolution is broken in two steps with an intermediate time, $t_f > t_m > t_0$, namely:
\begin{equation}
\ket{\rho(t_f)} = \Phi(t_f,t_m) \Phi(t_m,t_0)\ket{\rho(t_0)}. 
\end{equation} 
Whereas $\Phi(t_f,t_0)$ is a completely positive (CP) map for arbitrary $t_f$ \cite{Zyc06}, the map corresponding 
to the intermediate step, $\Phi(t_f,t_m)$,
 may  be non-CP for some $t_m$.
As realizable maps are always CP,   $\Phi(t_f,t_m)$ being non-CP  for the particular time $t_m$ witnesses the fact that 
such a division is not possible. A trivial case
in which any intermediate division is possible corresponds to unitary evolution. Markovian evolutions  also admit arbitrary
 intermediate steps. 
The intermediate map may fail to be CP only in the case of non-Markovian evolution.  This {\it divisibility criterion}
 \cite{Rivas10}   is 
therefore  a sufficient condition  to detect non-Markovianity. 

In order the check the complete positivity of a map, we use the well known duality between CP maps and positive operators, 
expressed by the Choi's theorem \cite{jam,Zyc06}.
First we define the unique dynamical matrix  associated to the map:
\begin{equation}\label{Choi}
D^{mn}_{\mu\nu}=\Phi^{m\mu}_{n\nu}=\bra{m\mu}\Phi\ket{n\nu},
\end{equation}
where Latin and Greek indices correspond to system and environment Hilbert spaces, respectively. The Choi's theorem 
states that the map ($\Phi$) is CP if and 
only if its dynamical matrix ($D$) is a positive semi-definite operator.  Finally, to check the complete positivity  of the 
intermediate map, we form the matrix of its super-operator by means of the product:
\begin{equation}\label{mapa-tm}
\Phi(t_f,t_m)=\Phi(t_f,t_0)\Phi^{-1}(t_m,t_0).
\end{equation}
Note that $\Phi(t,t_0)$ is the matrix representation of the map that evolves the system from the initial time $t_0$ to any time $t$.
$\Phi^{-1}(t_m,t_0)$ is the pseudo-inverse of $\Phi(t_m,t_0)$, and  thus evolves the system from $t_m$ to $t_0$. Therefore the matrix
product in Eq.\ref{mapa-tm} defines a matrix representation for the intermediate map. While the dynamical matrix $(D(t,t_0))$ corresponding to $\Phi(t,t_0)$ is always positive semi-definite, the one  $(D(t_f,t_m))$ related
to $\Phi(t_f,t_m)$ may happen to be non-positive, and in this case it witnesses a non-Markovian evolution.

\section{Dynamical Matrix for a General  Fermionic Quadratic Hamiltonian}
In the previous section, we reviewed  the formalism of quantum maps and the divisibility criterion.
We now   apply such  formalism to  environments described by general   fermionic quadratic Hamiltonians, 
 interacting with a qubit. We will show how to obtain the exact expression for the Kraus decomposition of the dynamical matrix.
   
Let us then consider a general fermionic quadratic Hamiltonian, namely, 
\begin{equation}\label{QH}
H_g = \sum_{m,n=1}^L (x_{m,n} \ad{m} a_n + y_{m,n} \ad{m} \ad{n} + h.c.).
\end{equation}
where $L$ is the lattice size, and $x_{m,n},y_{m,n}$ are arbitrary complex numbers.
$a^\dagger_j (a_j)$   is the creation  (annihilation) operator, 
satisfying the usual anti-commutation relations:
\begin{equation}
\{a_i,a_j^\dagger\}=\delta_{ij}, \,\, \{a_i, a_j\}=0.
\end{equation}
 
 For the interaction of the qubit with this environment, we consider the following Hamiltonian:
\begin{equation} 
\label{Hint}
H_{int} = -\delta \ketbra{e}{e}\otimes V_e,
\end{equation}
where $\ket{g}$ and  $\ket{e}$ are the qubit ground and excited states, respectively,   and $V_e$ is
a  fermionic quadratic Hamiltonian. 
We consider that the qubit and environment are initially uncorrelated, and they are in an arbitrary pure initial state,

\begin{equation}
\ket{\psi(0)}=\ket{\chi(0)} \otimes \ket{\varphi(0)}   =( c_g \ket{g} + c_e \ket{e} )\otimes
\ket{\varphi(0)}, 
\label{ket.psi.total.evolution}
\end{equation}
 where $\ket{\chi(0)} = c_g \ket{g} + c_e \ket{e}$, with $|c_g|^2 + |c_e|^2 = 1$, is the initial qubit state.
The evolution  under the total Hamiltonian,
 \begin{equation}\label{Ht}
 H= H_g + H_{int},
 \end{equation}
  is given by:
 \begin{equation}
 \ket{\psi(t)} =   e^{-iHt/\hbar}\ket{\chi(0)}\otimes\ket{\varphi(0)} , 
 \end{equation}
 \begin{equation}
   \ket{\psi(t)} = c_g \ket{g}\otimes \underbrace{e^{-i H_g t/\hbar}\ket{\varphi(0)}}_{\ket{\varphi_g(t)}} + 
c_e \ket{e}\otimes \underbrace{e^{-i H_e t/\hbar}\ket{\varphi(0)}}_{\ket{\varphi_e(t)}},
\label{psi.t.Htotal}
\end{equation}
where
\begin{equation}\label{He}
H_e = H_g -\delta V_e.
\end{equation}
Such Hamiltonians, $H_e$  and $H_g$, can be easily diagonalized by a  Bogoliubov transformation \cite{Sachdev},
namely:
\begin{eqnarray}
B_{\pm k} &\equiv& \cos\left(\frac{\theta_{g}^k}{2}\right) a_{\pm k}\mp i \sin\left(\frac{\theta_{g}^k}{2}\right)\ad{\mp k}, \label{bog.trans.quadractic1}\\
 A_{\pm k} &\equiv& \cos\left(\frac{\theta_{e}^k}{2}\right) a_{\pm k}\mp i \sin\left(\frac{\theta_{e}^k}{2}\right)\ad{\mp k} \label{bog.trans.quadractic2}.
\end{eqnarray}
  These  new fermionic operators are related according to 
 \begin{equation}
B_{\pm k} = \cos(\alpha_k) A_{\pm k} \mp i \sin(\alpha_k)A_{\mp k}^{\dagger},
\end{equation}
where $\alpha_k = (\theta_g^k - \theta_e^k)/2$. 
The Hamiltonians  in diagonal form read:
\begin{equation}
H_g = \sum_k \epsilon_{g}^k (B_k^{\dagger} B_k + C_g), \, \, H_e = \sum_k  \epsilon_{e}^k (A_k^{\dagger} A_k + C_e),
\end{equation}
where $C_g$ and $C_e$ are  both real  constants, and  $ \epsilon_{g(e)}^k $ are the 
single-particle eigenvalues.
 The ground states of $H_g$ ($G_g$) and $H_e$ ($G_e$) are related by: 
\begin{equation}
\ket{G_g} = \prod_{k>0} \left[\cos(\alpha_k) + i \sin(\alpha_k) A_k^{\dagger}A_{-k}^{\dagger}\right] \ket{G_e}.
\label{ground.states.relation}
\end{equation}


Now we derive  the  Kraus decomposition of the  map super-operator ($\Phi$). 
  The Kraus operators of the  evolution are:
\begin{equation}\label{kraus}
K_i = (\mathbb{I}_S \otimes \bra{i})\, e^{-iH t/\hbar}\, (\mathbb{I}_S\otimes \ket{\varphi(0)}),
\end{equation}
with $ \mathbb{I}_S=\ketbra{g}{g}+\ketbra{e}{e}$. 
 Assuming, without loss of generality
  (the map does not depend on the initial states of the qubit-environment),
 that the 
environment  is initially in its ground state ,
 $\ket{\varphi(0)}=\ket{G_g}$, and using Eq.(\ref{psi.t.Htotal}), we obtain:
  \begin{equation} \label{kraus operator}
  K_i   =  \mathbb{I}_S \otimes \bra{i}
   \left[ \ketbra{g}{g} \otimes \ket{\varphi_g(t)} + \ketbra{e}{e} \otimes \ket{\varphi_e(t)}\right].  
\end{equation}
The environment states $\ket{\varphi_g(t)}$ and $\ket{\varphi_e(t)}$ are given by:
\begin{eqnarray}\label{varphi.t.ground}
& \ket{\varphi_g(t)} = e^{-iH_g t/\hbar} \ket{G_g} = e^{-iE_g t/\hbar} \ket{G_g}=& \\ \nonumber
 & e^{-iE_g t/\hbar} \prod\limits_{k>0} \left[\cos(\alpha_{k}) + i \sin(\alpha_{k}) A_{k}^{\dagger}A_{-k}^{\dagger}\right] \ket{G_e},
\end{eqnarray}
where $E_g$ is the ground state energy of $H_g$.  Likewise, using 
 Eq.(\ref{ground.states.relation}), we obtain:
 \begin{eqnarray}\label{varphi.t.excited}
& \ket{\varphi_e(t)} = e^{-iH_e t/\hbar} \times & \\  \nonumber
&  \prod\limits_{k>0} \left[\cos(\alpha_{k}) +i \sin(\alpha_{k}) A_{k}^{\dagger}A_{-k}^{\dagger}\right] \ket{G_e}=& \\ \nonumber
& \prod\limits_{k>0} \left[\cos(\alpha_{k}) + e^{-i(\epsilon_e^{k} + \epsilon_e^{-k})t/\hbar} i \sin(\alpha_k) A_{k}^{\dagger}A_{-k}^{\dagger}\right] \times & \\ \nonumber
 & e^{-i E_e t/\hbar}\ket{G_e}.&
 \end{eqnarray}
  In order to obtain the Kraus operators, it is enough
 to calculate the overlaps $\braket{i}{\varphi_g(t)}$ and $\braket{i}{\varphi_e(t)}$, for a given environment basis $\{\ket{i}\}$, as
  shown in Eq.(\ref{kraus operator}). A convenient basis is formed by the eigenstates of $H_e$, namely:
\begin{eqnarray}
\{\ket{i}\} = \{\ket{G_e}, A^{\dagger}_{\vec{k}_N} \ket{G_e}\},
\end{eqnarray}
where $\vec{k}_N = (k_1,k_2,...,k_N)$ is the vector representing the momentum of the $N(=1, \ldots, L)$ excitations, and $A^{\dagger}_{\vec{k}}
  = A^{\dagger}_{k_1} A^{\dagger}_{k_2} ... A^{\dagger}_{k_N}$.
 It is easy to see that the only non null  elements for ``$\braket{i}{\varphi_g(t)}$'', using Eq.(\ref{varphi.t.ground}), are given by,
\begin{equation}
\label{prod1}
\braket{G_e}{\varphi_g(t)} = e^{-i E_g t/\hbar} (\prod\limits_{k>0} \cos(\alpha_{k})),
\end{equation}
and
\begin{eqnarray}
 & a_{\vec{k}_N}(t)\equiv\bra{G_e} A_{-\vec{k}_N} A_{\vec{k}_N} \ket{\varphi_g(t)} =  & \\ \nonumber
 & e^{-i E_g t/\hbar} \prod\limits_{k \in \vec{k}_N}(i\sin(\alpha_{k})) (\prod\limits_{k>0 , \,k \notin \vec{k}_N}\cos(\alpha_{k}) ),
\end{eqnarray}
where $N$ varies from 1 to $L/2$. Analogously, the non null terms for ``$\braket{i}{\varphi_e(t)}$'', using 
Eq.(\ref{varphi.t.excited}), are given by,
\begin{eqnarray}
 &b_{\vec{k}_N}(t)\equiv \bra{G_e} A_{-\vec{k}_N} A_{\vec{k}_N}  \ket{\varphi_e(t)} =  & \\ \nonumber
 & e^{-i E_e t/\hbar} \prod\limits_{k \in \vec{k}_N} \left[ i\sin(\alpha_{k})\exp( -i(\epsilon_e^{k}+\epsilon_e^{-k}) t/\hbar )\right] \times & \\ \nonumber
  & (\prod\limits_{k>0, \,k \notin \vec{k}_N} \cos(\alpha_{k}) ).
\end{eqnarray}

It is easy to check the following relation:
\begin{equation}
 b_{\vec{k}_N}(t) =a_{\vec{k}_N}(t)f_{\vec{k}_N}(t),
\end{equation}
where
\begin{equation}
f_{\vec{k}_N}(t)\equiv e^{-i(E_e - E_g)t/\hbar}\exp(-i\sum_{k \in \vec{k}_N}^N (\epsilon_e^{k}+\epsilon_e^{-k})t/\hbar).
\end{equation}
Finally,  we reach the first result of this work, obtaining a simple expression for the Kraus operators of the quantum map,
\begin{equation}
K_{\vec{k}_N} =
 a_{\vec{k}_N}(t)(\ketbra{g}{g} + f_{\vec{k}_N}(t) \ketbra{e}{e}).
\end{equation}
Note that  $\abs{a_{\vec{k}_N}(t)}^2$ is not a time dependent variable, and 
 \begin{equation}
 \sum_{\{\vec{k}_N\}} \abs{a_{\vec{k}_N}(t)}^2 = Tr(\ketbra{\varphi_g(t)}{\varphi_g(t)}) = 1.
 \end{equation}
By using this fact, we can then write the quantum map in terms of the Kraus operators as follows,
\begin{eqnarray}
\Phi(t,0) &=& \sum_{\{{\vec{k}_N}\}}K_{\vec{k}_N} \otimes K_{\vec{k}_N}^* \nonumber\\
&= & \ketbra{g}{g}
 \otimes \ketbra{g}{g} + \ketbra{e}{e} \otimes \ketbra{e}{e}+ \nonumber\\ 
& &  \ketbra{g}{g} \otimes \ketbra{e}{e} \sum_{\{{\vec{k}_N}\}}\abs{a_{\vec{k}_N}(t)}^2 f_{\vec{k}_N}(t)^*+ \\
& &  \ketbra{e}{e} \otimes \ketbra{g}{g} \sum_{\{{\vec{k}_N}\}}\abs{a_{\vec{k}_N}(t)}^2 f_{\vec{k}_N}(t)\nonumber
\end{eqnarray}
If we define the following variable,
\begin{equation}
x(t)\equiv\sum_{\{\vec{k}_N\}} \abs{a_{\vec{k}_N}(t)}^2 f_{\vec{k}_N}(t),
\end{equation}
the quantum map can be rewritten as,
\begin{eqnarray}\label{phit0}
& \Phi(t,0) = \,\left[ \ketbra{g}{g} \otimes \ketbra{g}{g} + \ketbra{e}{e} \otimes \ketbra{e}{e} + \right. & \nonumber\\ 
&  \left. \ketbra{g}{g} \otimes \ketbra{e}{e} x(t)^* + \ketbra{e}{e} \otimes \ketbra{g}{g} x(t) \right].   \label{mapa.Phi}
\end{eqnarray}

As expected, the quantum map consists in a decoherence channel,
 and  thus we can identify the variable ``$x(t)$'' with the known Loschmidt echo $\mathcal{ L}( t  )$ \cite{Quan06, Haikka12},

\begin{equation}
 \mathcal{L}(t) = |x(t)|^2 = |\braket{\phi_g(t)}{\phi_e(t)}|^2.
\label{relation.x.loschmidt}
\end{equation}
The above relation follows just by noticing that the qubit reduced state, 
$\rho_S(t) = Tr_E(\ketbra{\psi(t)}{\psi(t)})$, taking the 
partial trace of Eq.\eqref{ket.psi.total.evolution}, is given by
 $\rho_S(t) = |c_g|^2 \ketbra{g}{g} + |c_e|^2 \ketbra{e}{e} + 
 c_g^* c_e \mu(t) \ketbra{e}{g} + H.c. $, where $\mu(t)=\braket{\phi_g(t)}{\phi_e(t)}$
  is the decoherence factor. The quantum map corresponding to such an evolution is 
  the decoherence channel, as described before.

Using now Eq.(\ref{mapa-tm}), we have the following expression for the intermediate map:
\begin{eqnarray}
 \Phi(t_f,t_m) = \left[ \ketbra{g}{g} \otimes \ketbra{g}{g} + \ketbra{e}{e} \otimes \ketbra{e}{e} + \right.   \nonumber\\ 
   \ketbra{g}{g} \otimes \ketbra{e}{e} y(t_f,t_m)^* +    \\ \nonumber
   \left. \ketbra{e}{e} \otimes \ketbra{g}{g} y(t_f,t_m)\label{mapa.Phi2}\right],
\end{eqnarray}
where
\begin{equation}
y(t_f,t_m)\equiv \frac{x(t_f)}{x(t_m)}.
\end{equation}
The dynamical matrix of this quantum map is
\begin{equation}
D_{\Phi(t_f,t_m)} = \left(\begin{matrix}
		1 & 0 & 0 & y(t_f,t_m)^* \\\\
		0 & 0 & 0 & 0 \\\\
		0 & 0 & 0 & 0 \\\\
		y(t_f,t_m) & 0 & 0 & 1 \\\\
\end{matrix}\right).
\end{equation}
Computing the minimum eigenvalue, we arrive at the following simple sufficient condition for the positive-semi-definiteness of the dynamical matrix: 
\begin{equation}
\label{dynamicalH} 
1 - |y(t_f,t_m)| \geq 0.
\end{equation}
Therefore we have obtained a simple  function capable to witness the non-Markovianity of the dynamics,
{\it i.e.}, $\Phi$ is non-Markovian if $|y(t_f,t_m)|>1$.

\section{Ising  model as an  environment for a  system of one qubit }
\begin{figure}
\includegraphics[scale=0.22]{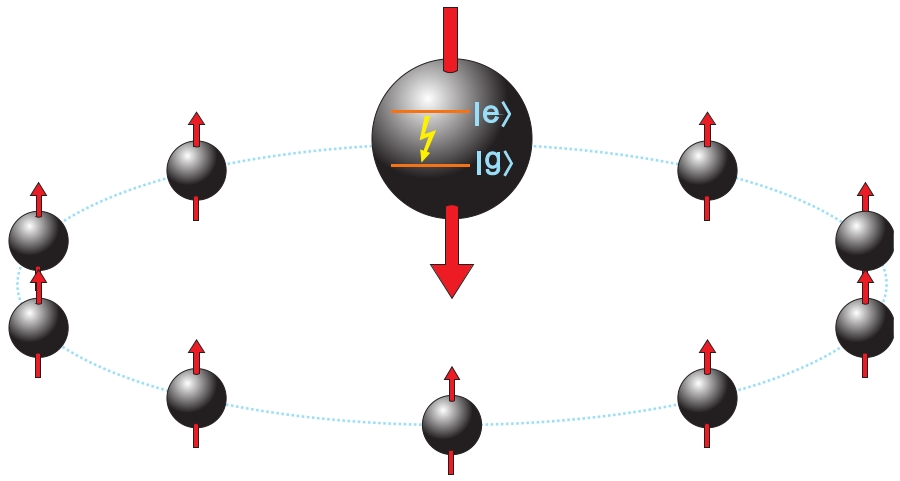}
\caption{ Schematic view of spins forming a ring array, representing 
 the environment  governed by the Ising Hamiltonian (Eq.\eqref{Hising}). The central spin is  the qubit interacting with 
the environment according to Eq.\eqref{Hint.X}. } 
\label{ising}
\end{figure}

In the previous section, we derived the dynamical matrix for an arbitrary quadratic fermionic Hamiltonian.
In this section we focus on an environment described by  the Ising Hamiltonian 
in a transverse field ($H_{ising}$), with periodic boundary conditions ($L+1=1$). 
The  interaction with the environment ($H_{int}$) is by means of  the transverse  magnetic
  field in the $Z$ direction (see Fig.\ref{ising}), more precisely,
\begin{eqnarray}
H_{Ising} &=& -J \sum_{j=1}^L ({\sigma^x_j}{\sigma^x_{j+1}} + \lambda {\sigma^z_j}), \label{Hising}\\
H_{int} &=& -\delta \ketbra{e}{e}\otimes\sum_{j =1}^L{\sigma^z_j}. \label{Hint.X}
\end{eqnarray}

In order to employ the previous section's results, we first do the identification:
\begin{eqnarray}
    H_e &=& H_{ising} - \delta 
   \sum_{j =1}^L{\sigma^z_j},   \\ 
  H_g &=& H_{ising}. 
 \end{eqnarray}

We now diagonalize the Ising Hamiltonian \cite{Franchini}. First we use the usual Jordan-Wigner transformation,
\begin{eqnarray}\label{JW}
\sigma_j^+ &=&  \exp{(i\pi \sum_{l<j} a_l^\dagger a_l)}  =\prod_{l<j} (1 - 2a_l^\dagger a_l) a_j, \\
a_j&=&(\prod_{l<j} \sigma_l^z)\sigma_j^+.
\end{eqnarray}
The Ising Hamiltonian can then be rewritten in terms of quadratic  fermionic operators:
\begin{eqnarray}
H_{ising} = J \left[-\sum\limits_{j=1}^{L-1}(\ad{j}a_{j+1} + \ad{j}\ad{j+1} + h.c.)  \right. \nonumber \\
\left. + e^{(i\pi)\hat{N}} (\ad{L}a_1 + \ad{L}\ad{1} + h.c.) + 2\lambda \hat{N}  - \lambda L \right],
\end{eqnarray}
where $\hat{N} = \sum_j \ad{j}a_j$. The Hamiltonian conserves the parity, $\comm{H}{e^{(i\pi)\hat{N}}} = 0$. 
Thus we can analyze its $odd/even$ subspaces separately. The gap between the ground state energy of these two subspaces
 obviously closes in the thermodynamic limit. For simplicity, we shall proceed the analysis in the $even$ sector, which leads to a simple
  quadratic Hamiltonian with anti-periodic boundary conditions. Using the momentum eigenstates,
\begin{equation}
a_{k} = \frac{1}{\sqrt{L}} \sum_j e^{(-i k j)} a_{j},
\end{equation}
with $k = \frac{2\pi}{L}q,\,\, q=\pm1/2,\pm3/2,...,\pm(L-1)/2$, for $L$ even, and the Bogoliubov transformation  (Eq.(\ref{bog.trans.quadractic2})), with phases
 \begin{equation}
\theta_e^k(\delta) = \arctan\left[\frac{-\sin(k)}{\cos(k)-(\lambda+\delta)}\right],
\end{equation}
the Hamiltonian assumes the  desired diagonal form:
\begin{equation}\label{H.ising.diag}
H_e = \sum_k \epsilon_e^k \,(A_k^{\dagger}A_k - 1/2),
\end{equation}
with eigenvalues given by:
\begin{equation}
\epsilon_e^k(\delta) = J\sqrt{1 + (\lambda + \delta)^2 - 2(\lambda + \delta)\cos(k)}.
\end{equation}

\section{Ising model as an environment for a system of  two qubits}

Now we determine  the exact expression for the quantum map$(\Phi)$, in  the case of two qubits interacting with an environment described by an
arbitrary  quadratic fermionic Hamiltonian $H_{g}$ (Eq.(\ref{QH})). The motivation is to investigate how the number of particles in the system affects
the environment.

We assume the two qubits described by the Hamiltonian
\begin{equation}
H_S=-J_S\left[\sigma_{1}^{z}\sigma_{2}^{z}+\lambda_S\left(\sigma_{1}^{z}+\sigma_{2}^{z}\right)\right],
\end{equation}
where $\sigma^{z}=\ketbra{g}{g}-\ketbra{e}{e}$, with $\ket{g}$ and  $\ket{e}$ being the qubit ground and excited states. For the interaction with the environment, we consider the following Hamiltonian:
\begin{eqnarray}
H_{int}&=&-\left[ \delta_1 \ketbra{gg}{gg}+\delta_2\left( \ketbra{ge}{ge} \right. \right. \nonumber\\
&& \left. \left.+\ketbra{eg}{eg}\right)\right]\otimes V,
\end{eqnarray}
where $V$ is a fermionic quadratic Hamiltonian. We assume that the two qubits and the environment are initially uncorrelated, and they are in an arbitrary pure initial state,
\begin{equation}
  |\psi(0)\rangle=|\chi(0)\rangle\otimes  |\varphi(0)\rangle,\nonumber \\
\end{equation}
where $|\chi(0)\rangle=c_{gg}\ket{gg}+c_{ge}\ket{ge}+c_{eg}\ket{eg}+c_{ee}\ket{ee}$  
( $|c_{gg}|^2+|c_{ge}|^2+|c_{eg}|^2+|c_{ee}|^2=1$) is the initial two-qubit state. Therefore, the state of the composite system, at an arbitrary time $t$, can be written as:
\begin{eqnarray}\label{is}
  |\Psi(t)\rangle&=&e^{-i\left(H_g+H_{int}+H_S\right)t/\hbar}\ket{\chi(0)}\otimes\ket{\varphi(0)}\nonumber\\\nonumber\\
&=&e^{-iJ_St/\hslash}\left(c_{ge}\ket{ge}+c_{eg}\ket{eg}\right)\ket{\varphi_2(t)}+\nonumber\\
&&c_{gg}e^{iJ_S(1+2\lambda)t/\hslash}\ket{gg}\ket{\varphi_1(t)}+\nonumber\\
&&c_{ee}e^{iJ_S(1-2\lambda)t/\hslash}\ket{ee}\ket{\varphi_0(t)},
\end{eqnarray}
where $|\varphi_a(t)\rangle= e^{-iH_at/\hslash}|\varphi(0)\rangle$, with Hamiltonian $H_a=H_0-\delta_aV_e$, $a=(0,1,2)$, and $\delta_0=0$. With this notation, we have $H_0\equiv H_g$.

The Hamiltonian $H_a$$(a=(0,1,2))$ can be diagonalized by a Bogoliubov transformation, 
\begin{equation}\label{ferm}
\eta_a^{\pm k} =\cos\left(\frac{\theta_a^k}{2}\right) a_{\pm k}\mp i\sin\left(\frac{\theta_a^k}{2}\right) a_{\mp k}^\dagger.
\end{equation}
These fermionic operators are related by:
\begin{equation}\label{rel.ferm}
\eta_a^{\pm k} =\cos\left(\alpha_{a,b}^k\right) \eta_b^{\pm k}\mp i\sin\left(\alpha_{a,b}^k\right)\eta_b^{\mp k\dagger},
\end{equation}
where $\alpha_{a,b}^k=(\theta_a^k-\theta_b^k)/2$. The Hamiltonian in diagonal form reads:
\begin{equation}\label{hamilt}
H_a=\sum_k\epsilon_a^{k}\left(\eta_a^{k\dagger}\eta_a^{k}+C_a\right),
\end{equation}
where $C_0$, $C_1$ and $C_2$ are real constants, $\epsilon_0^{k}$, $\epsilon_1^{k}$ and $\epsilon_2^{k}$ are the single-particle eigenvalues. The ground states of $H_0$ $(G_0)$, $H_1$ $(G_1)$ and $H_2$ $(G_2)$ are related according to:
\begin{equation}\label{lis}
|G_a\rangle=\prod_{k>0}\left[\cos\left(\alpha_{a,b}^k\right)+i\sin\left(\alpha_{a,b}^{k}\right) \eta_{b}^{k\dagger}\eta_{b}^{-k\dagger}\right]|G_b\rangle.
\end{equation}

Using the definition of Kraus operators in Eq.(\ref{kraus}),  and the Eq.(\ref{is}), we can write:
\begin{eqnarray}
K_{i}&=&\braket{i}{\varphi_2}e^{-iJt/\hslash}\left(\ketbra{ge}{ge}+\ketbra{eg}{eg}\right)\nonumber\\
&&\braket{i}{\varphi_1}e^{iJ(1+2\lambda)t/\hslash}\ketbra{gg}{gg}+\nonumber\\
&&\braket{i}{\varphi_0}e^{iJ(1-2\lambda)t/\hslash}\ketbra{ee}{ee},
\end{eqnarray}
where $\left\{\ket{i}\right\}$ is an environment basis. Finally we obtain the quantum map:
\begin{eqnarray}
\Phi\left(t,0\right)&=&\sum_{i}K_{i}\otimes K_{i}^{*}\nonumber \\
&=&\left[\ketbra{gg}{gg}\otimes\ketbra{gg}{gg}+\ketbra{ee}{ee}\otimes\ketbra{ee}{ee}+\right.\nonumber \\
&&\ketbra{ge}{ge}\otimes\ketbra{ge}{ge}+\ketbra{eg}{eg}\otimes\ketbra{eg}{eg}+\nonumber \\
&&\ketbra{ge}{ge}\otimes\ketbra{eg}{eg}+\ketbra{eg}{eg}\otimes\ketbra{ge}{ge}+\nonumber \\
&&\left(\ketbra{ee}{ee}\otimes\ketbra{eg}{eg}+\ketbra{ee}{ee}\otimes\ketbra{ge}{ge}\right)\times\nonumber \\
&&x_{0,2}(t)^{*}e^{i\phi_{-}t}+\nonumber \\
&&\left(\ketbra{eg}{eg}\otimes\ketbra{ee}{ee}+\ketbra{ge}{ge}\otimes\ketbra{ee}{ee}\right)\times\nonumber \\
&&x_{0,2}(t)e^{-i\phi_{-}t}+\nonumber \\
&&\left(\ketbra{gg}{gg}\otimes\ketbra{eg}{eg}+\ketbra{gg}{gg}\otimes\ketbra{ge}{ge}\right)\nonumber \\
&&x_{1,2}(t)^{*}e^{i\phi_{+}t}+\nonumber \\
&&\left(\ketbra{eg}{eg}\otimes\ketbra{gg}{gg}+\ketbra{ge}{ge}\otimes\ketbra{gg}{gg}\right)\nonumber \\
&&x_{1,2}(t)e^{-i\phi_{+}t}+\nonumber \\
&&\ketbra{ee}{ee}\otimes\ketbra{gg}{gg}x_{0,1}(t)^{*}e^{-i\phi_{0}t}+\nonumber \\
&&\left.\ketbra{gg}{gg}\otimes\ketbra{ee}{ee}x_{0,1}(t)e^{i\phi_{0}t}\right],\nonumber \\
\end{eqnarray} 
with, $\phi_{\pm}=2J_S(1\pm\lambda_S)/\hslash$ and $\phi_{0}=4J_S\lambda_S/\hslash$, and $x_{a,b}(t)=\braket{ \varphi_b}{ \varphi_a}$.
Choosing the environment  in its initial ground state, $\ket{\varphi(0)}=\ket{G_0}$,  and using equations (\ref{rel.ferm}-\ref{lis}), we have:
\begin{eqnarray}
x_{a,b}(t)&=&\langle \varphi_b| \varphi_a\rangle\nonumber\\
&=&\prod_{k>0}\left\{\cos\left(\alpha_{0,a}^{k}\right)\cos\left(\alpha_{0,b}^{k}\right)\cos\left(\alpha_{a,b}^{k}\right)+\right.\nonumber\\
&&\left[\cos\left(\alpha_{0,a}^{k}\right)\sin\left(\alpha_{0,b}^{k}\right)e^{i(\epsilon_b^k+\epsilon_b^{-k})t/\hbar}\right. -\nonumber\\
&&\left.\cos\left(\alpha_{0,b}^{k}\right)\sin\left(\alpha_{0,a}^{k}\right)e^{-i(\epsilon_a^k+\epsilon_a^{-k})t/\hbar}\right]\times\nonumber\\
&&\sin\left(\alpha_{a,b}^{k}\right)+\sin\left(\alpha_{0,a}^k\right)\sin\left(\alpha_{0,b}^k\right)\cos\left(\alpha_{a,b}^{k}\right)\times\nonumber\\
&& \left.e^{-i\left[(\epsilon_a^k+\epsilon_a^{-k})-(\epsilon_b^k+\epsilon_b^{-k})\right]t/\hbar}\right\}e^{-i(E_a-E_b)t/\hbar},
\end{eqnarray}
where $E_a$ is the ground state energy of $H_a$. Finally, we obtain the dynamical matrix of the  intermediate map, namely:

\begin{eqnarray}
D_{\Phi(t_f,t_m)}&=&\left[\ketbra{gg}{gg}\otimes\ketbra{gg}{gg}+\ketbra{ee}{ee}\otimes\ketbra{ee}{ee}+\right.\nonumber \\
&&\ketbra{ge}{ge}\otimes\ketbra{ge}{ge}+\ketbra{eg}{eg}\otimes\ketbra{eg}{eg}+\nonumber \\
&&\ketbra{ge}{eg}\otimes\ketbra{ge}{eg}+\ketbra{eg}{ge}\otimes\ketbra{eg}{ge}+\nonumber \\
&&\left(\ketbra{ee}{eg}\otimes\ketbra{ee}{eg}+\ketbra{ee}{ge}\otimes\ketbra{ee}{ge}\right)\times\nonumber \\
&&y_{0,2}(t_f,t_m)^{*}e^{i\phi_{-}(t_f-t_m)}+\nonumber \\
&&\left(\ketbra{eg}{ee}\otimes\ketbra{eg}{ee}+\ketbra{ge}{ee}\otimes\ketbra{ge}{ee}\right)\times\nonumber \\
&&y_{0,2}(t_f,t_m)e^{-i\phi_{-}(t_f-t_m)}+\nonumber \\
&&\left(\ketbra{gg}{eg}\otimes\ketbra{gg}{eg}+\ketbra{gg}{ge}\otimes\ketbra{gg}{ge}\right)\times\nonumber \\
&&y_{1,2}(t_f,t_m)^{*}e^{i\phi_{+}(t_f-t_m)}+\nonumber \\
&&\left(\ketbra{eg}{gg}\otimes\ketbra{eg}{gg}+\ketbra{ge}{gg}\otimes\ketbra{ge}{gg}\right)\times\nonumber \\
&&y_{1,2}(t_f,t_m)e^{-i\phi_{+}(t_f-t_m)}+\nonumber \\
&&\ketbra{ee}{gg}\otimes\ketbra{ee}{gg}y_{0,1}(t_f,t_m)^{*}e^{-i\phi_{0}(t_f-t_m)}+\nonumber \\
&&\left.\ketbra{gg}{ee}\otimes\ketbra{gg}{ee}y_{0,1}(t_f,t_m)e^{i\phi_{0}(t_f-t_m)}\right],\nonumber \\
\end{eqnarray}
where
\begin{equation}
y_{a,b}(t_f,t_m)=\frac{x_{a,b}(t_f)}{x_{a,b}(t_m)}.
\end{equation}

Unlike the case of one qubit, where we presented a very simple expression for the minimum eigenvalue of the dynamical matrix (Eq.(\ref{dynamicalH})), directly related to the well know Loschmidt echo, in the case of two qubits the minimum eigenvalue is a non-trivial function of the parameters $y_{a,b}(t_f,t_m)$. However, working numerically we learn that the two-qubit case does not present any new characteristic that would result in a different behavior of the non-Markovianity in relation to the one-qubit case.

\section{Witnessing  the non-Markovianity in the Ising Model: Finite size effects}

Now we are equipped to characterize the dynamics of a qubit 
interacting with an environment governed by the Ising model (Fig. 1).
We consider  lattices up to $L=5\times 10^{5}$
  sites, and investigate the non-Markovianity in 
  the vicinity of the critical point of the quantum Ising model, which  is 
  well known to be equal to $\lambda^{*}\equiv\lambda + \delta = 1$.

\begin{figure}[H]
\includegraphics[scale=0.57]{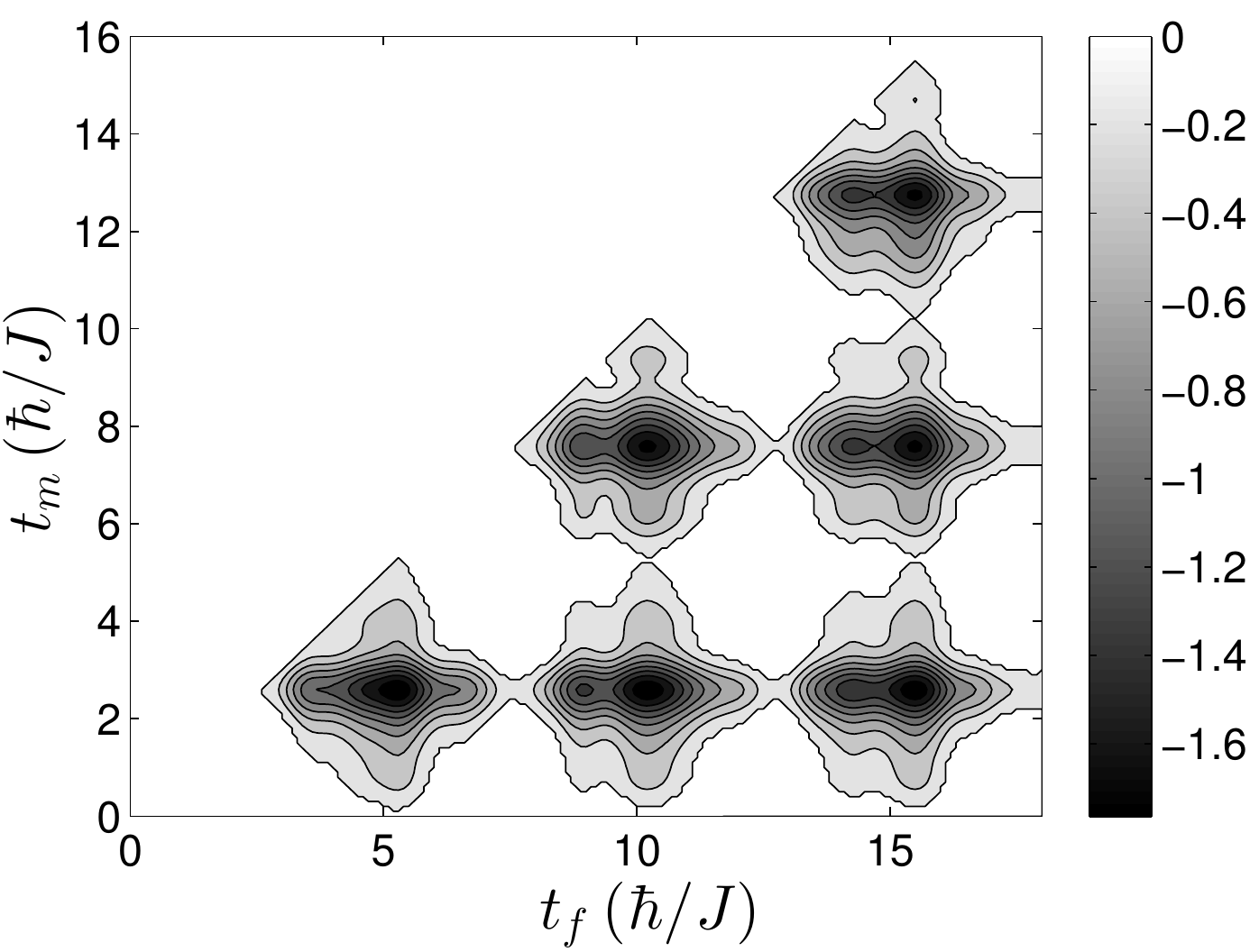}
 \caption{Manifestation of the non-Markovianity by means of the most negative eigenvalue of the intermediate quantum map $D_{\Phi(t_f,t_m)}$ (greyscale), in function of 
$t_f$ and $t_m$, for a lattice with parameters   $L=10$ , $\lambda=0.5$ and $\delta=0.5$.
\label{tf_tm}}
\end{figure}

Let us define a measure ($\eta$) of  non-Markovianity 
as the minimum of the eigenvalues for the intermediate quantum
 dynamical matrix $D_{\Phi(t_t,t_m)}$ over all 
 final times $t_f$ and over all time partitions $t_m$, precisely:
\begin{equation}
\eta=\min\limits_{\{t_f\}}\min\limits_{\{t_m<t_f\}} eig\{D_{\Phi(t_t,t_m)}\},
\label{measureN}
\end{equation}
where $eig$ is the set of eigenvalues 
of the intermediate dynamical matrix $D_{\Phi(t_f,t_m)}$.
In order to exemplify such 
a non-Markovianity measure, we plot, in Fig. 2, the 
smallest eigenvalue of the intermediate 
map  as a function of the final ($t_f$)  and 
intermediate  ($t_m$) times, at the critical point of the Ising
model, for a lattice with $L=10$ sites. As the values of
$t_m$ and $t_f$ are swept, the non-Markovian regions of the 
dynamics are revealed. 

 Notice that the previously defined non-Markovianity measure 
is only based on the minimum eigenvalue of the dynamical matrix. 
 One might expect, however, that the number of negative eigenvalues could
influence the strength of the non-Markovianity.
 For our models under analysis, however, it seems not play any 
 relevant effect: i) in the case of a single qubit it
becomes trivial, since one can only have a single negative eigenvalue for
 the dynamical matrix; ii) and in the case of two-qubits we found that
indeed there are cases where the dynamical matrix presents more than one 
negative eigenvalue, but its absolute
value is always at least two orders of magnitude smaller than the absolute 
value of the minimum eigenvalue,
and thus could be neglected.

 \begin{figure}[H]
\includegraphics[scale=0.4]{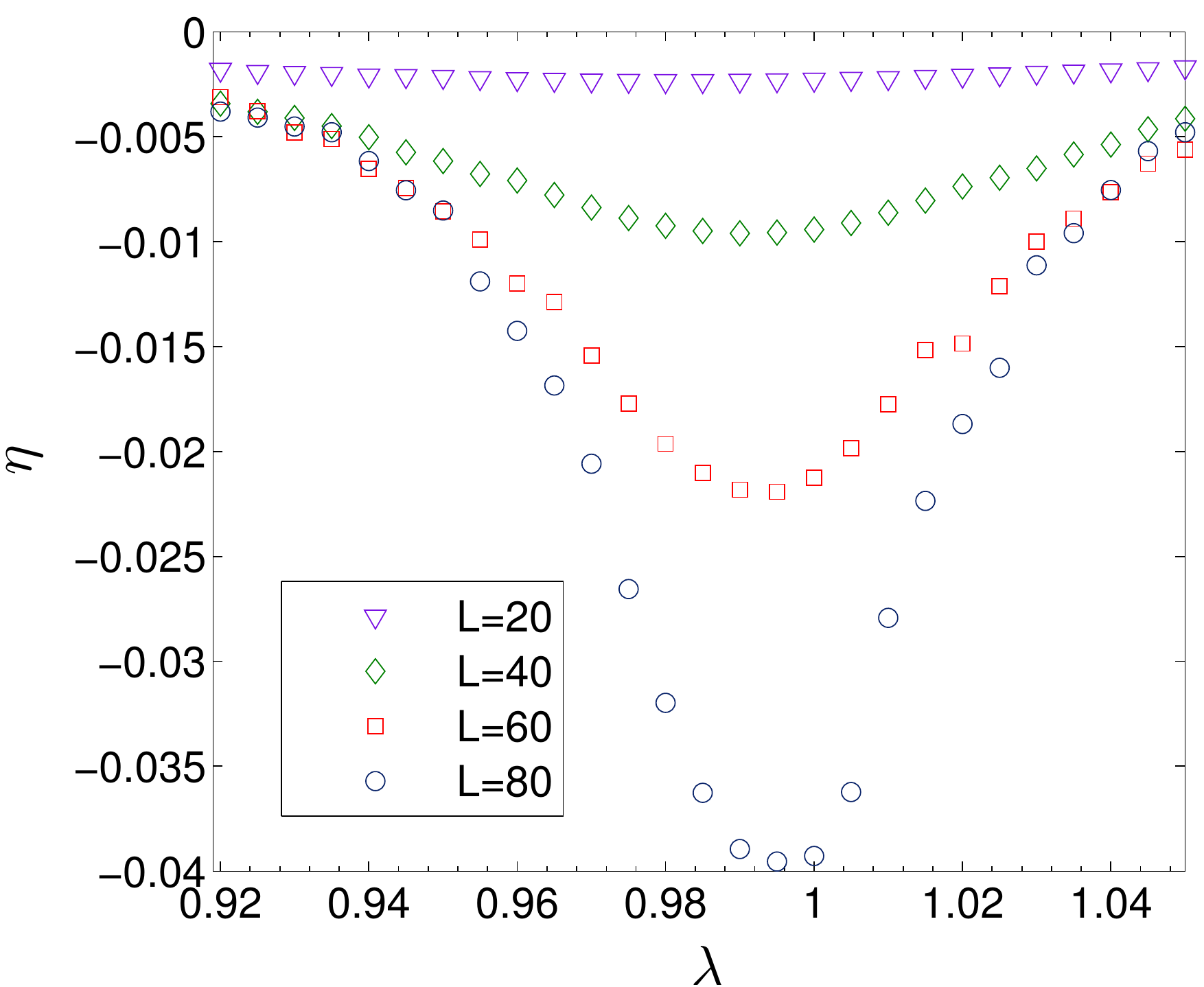}
\caption{The non-Markovianity  measure $\eta$ (Eq. (\ref{measureN})) in function of the transverse field $\lambda$, for $\delta=0.01$, and for different 
lattice sizes ($L$), in the vicinity of the Ising model  critical point.
\label{L20L80}}
\end{figure} 

In Fig. 3, the non-Markovianity, quantified by $\eta$ (Eq.(\ref{measureN})),  is plotted
 against the transverse field ($\lambda$), in the vicinity of
the Ising model critical point, for a fixed interaction coupling
 constant $\delta=0.01$. We see that the larger the lattice, the larger the  non-Markovianity.
The most interesting feature shown in this figure is
 the maximum of non-Markovianity occurring precisely at the
  Ising model critical point. 
The behavior of this measure for larger lattice sizes, 
and in the thermodynamic limit, for the particular model studied in this section could also be inferred
 by the Loschmidt echo \cite{Quan06,Haikka12},
from Eqs.\eqref{relation.x.loschmidt} and \eqref{dynamicalH}.
Note, however, that this equivalence
between $\eta$ and the Loschmidt echo is not necessarily true in  general.

\begin{figure}[H]
\includegraphics[width=8.0cm]{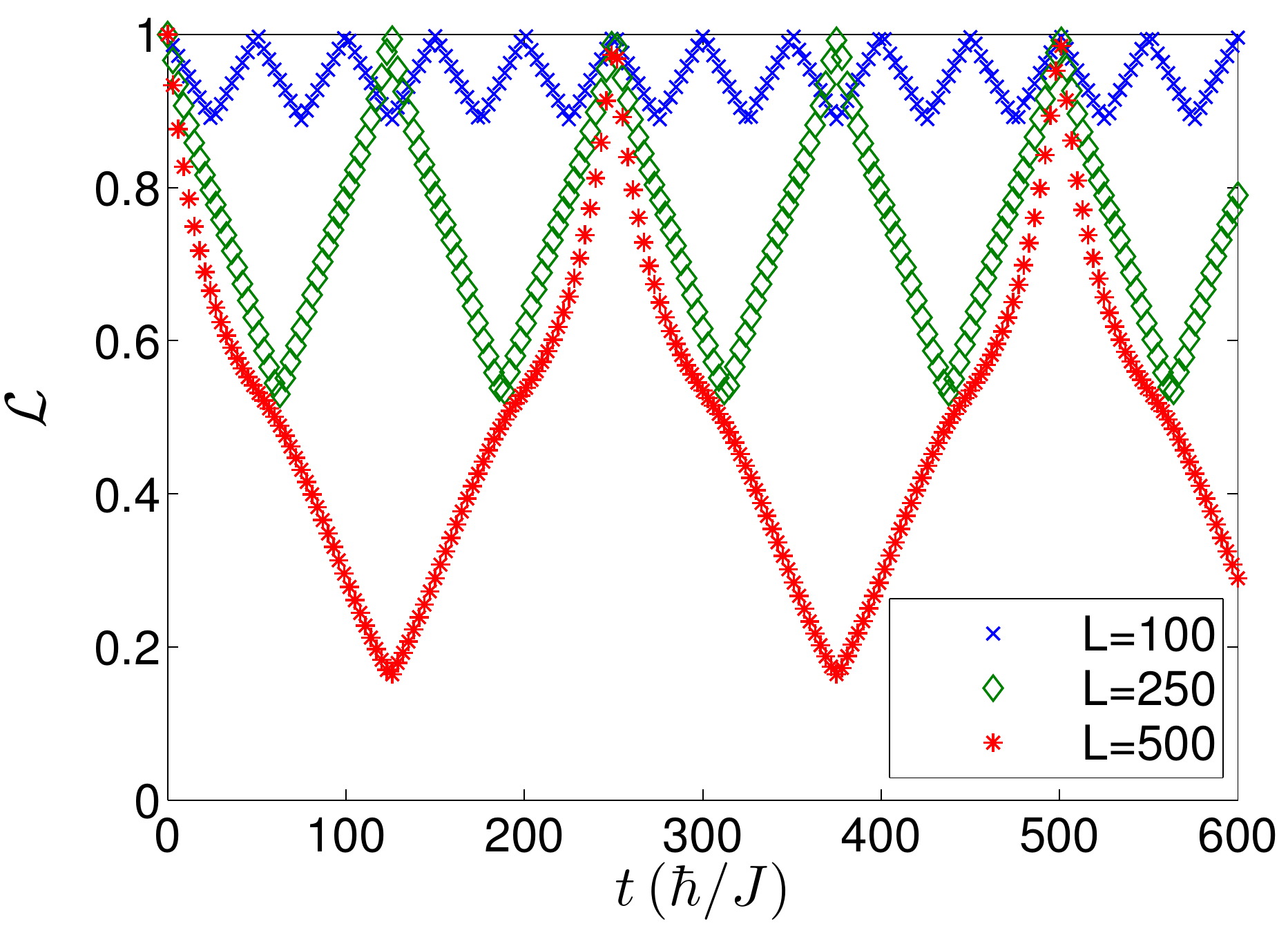}
  \caption{The Loschmidt echo $\mathcal{L}$ (Eq. (37)) as a function of the time, at the critical point $\lambda^*= \lambda + \delta = 1$, with $\delta=10^{-2}$,
  for different lattice sizes.
\label{fig.losch.cp} }
\end{figure}

 In Fig. \ref{fig.losch.cp} ,we see the behavior 
of the Loschmidt echo, for different lattice sizes, at the critical 
point ($\lambda^*=1$).
 We highlight some of its features:
 (i) the Loschmidt echo has an abrupt 
 decay followed by a revival, with a time period ``$\tau$'', which is
  proportional to the lattice size, $\tau \propto L$; (ii)
   the difference between the minimum value of the decay (which we shall 
   denote by $\mathcal{L}_{dec}$) and 
 the maximal of the revival ($\mathcal{L}_{rev}$) becomes higher as we increase the lattice size. 
 In this way, the non-Markovianity measure is simply 
 given by $\eta = {\mathcal L}_{rev} / {\mathcal L}_{dec}$.

 \begin{figure}[H]
\includegraphics[width=8.0cm]{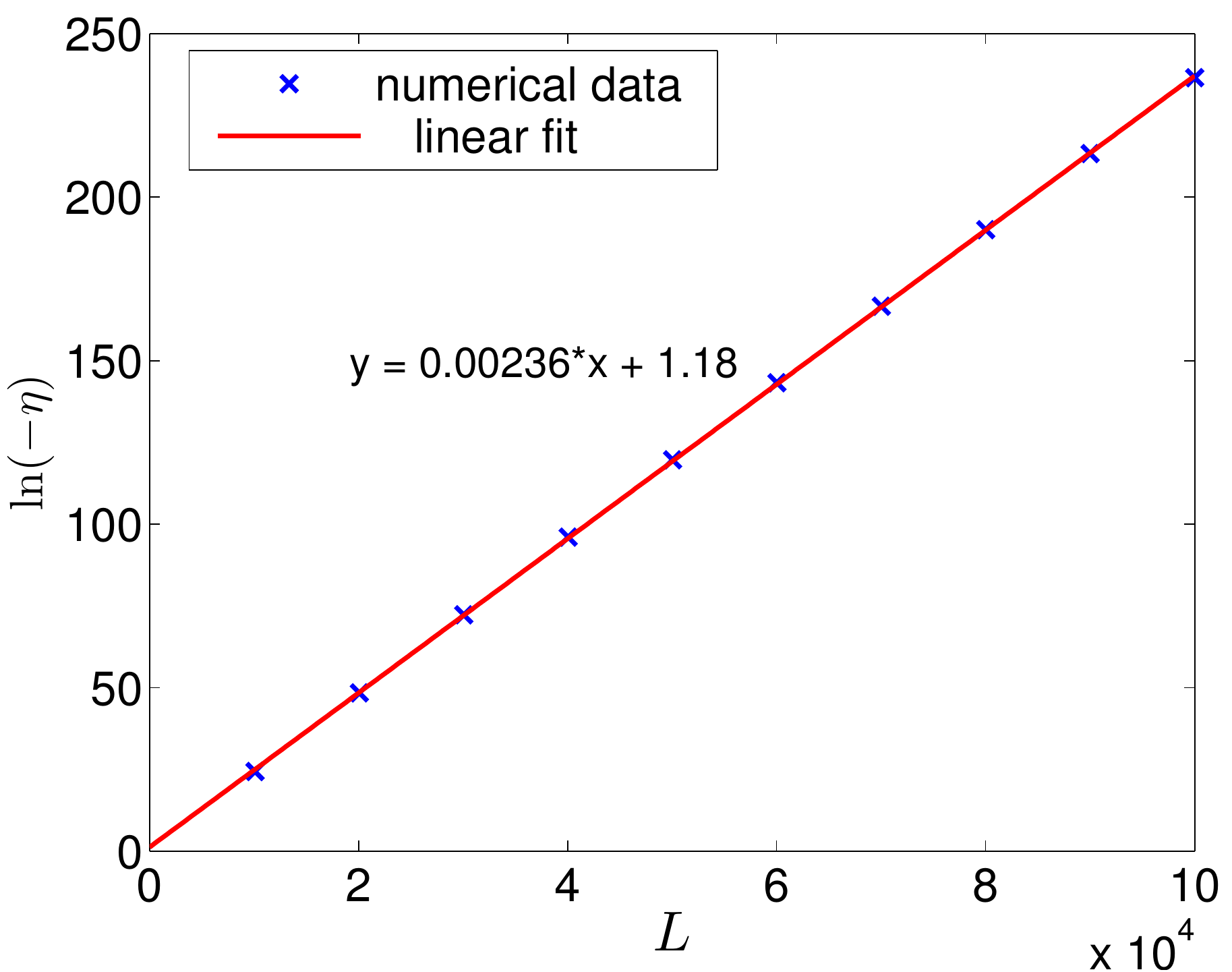}
  \caption{Finite size scaling analysis: $ln(-\eta)$ as a function 
of $L$, for $L=100$ to $L=10^5$ sites, at the critical 
point $\lambda^{*}=1$, with $\delta=10^{-2}$. The 
linear fit reveals an exponential
divergence of the non-Markovianity with the lattice size.
\label{fig.Nscaling.cp} }
\end{figure}

\begin{figure}[H]
\includegraphics[width=8.9cm]{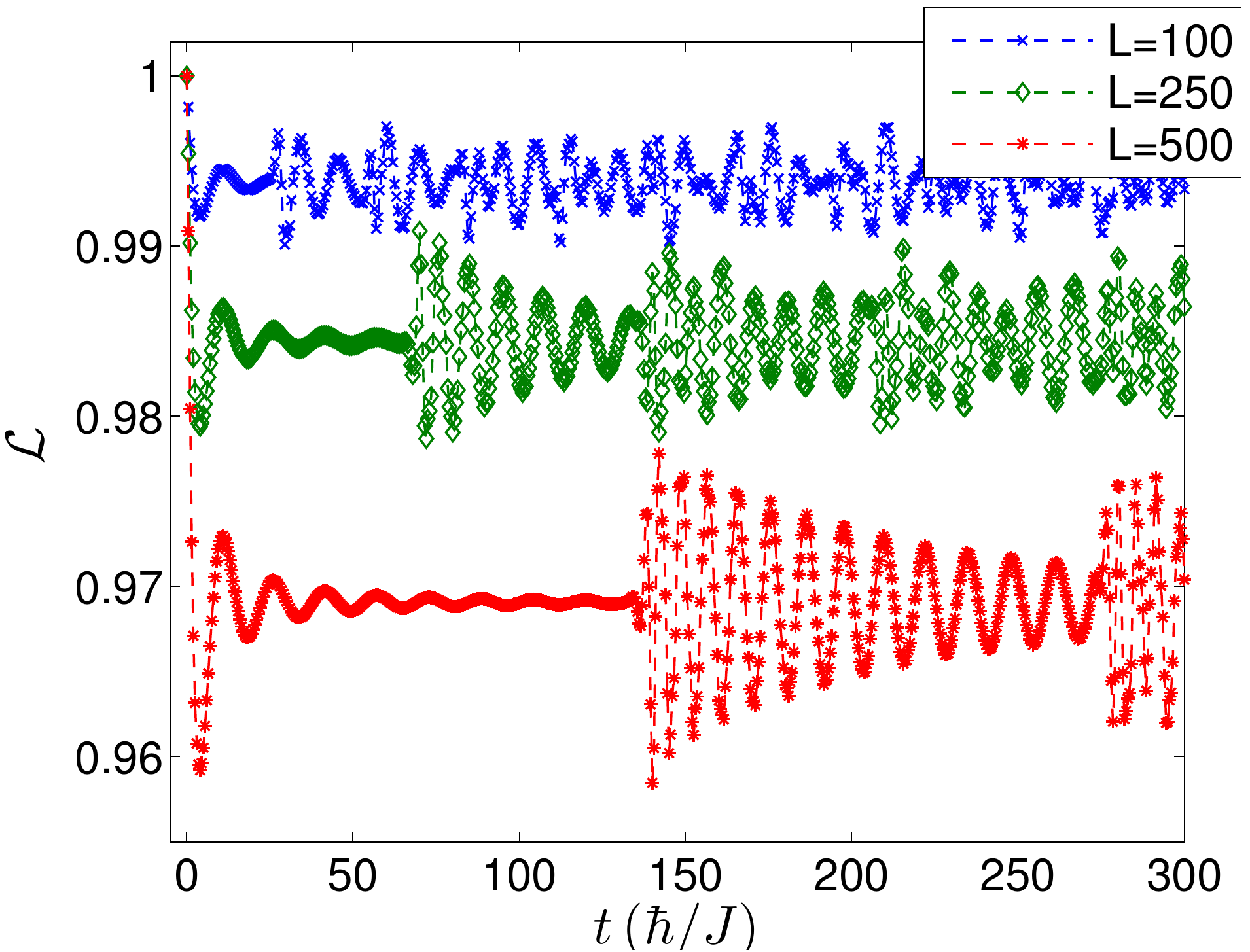}
 \caption{The Loschmidt echo  $\mathcal{L}$ (Eq. (37)) as a function 
of the time, outside of   the critical point; more precisely,
 for $\lambda = \lambda^* - 0.1 $, and $\delta=10^{-2}$.
  The behavior for $\lambda = \lambda^* + 0.1 $ is completely similar to this one.
\label{fig.losch.ncp} }
\end{figure}

 Performing a 
 finite-size scaling analysis, we see, in Fig. \ref{fig.Nscaling.cp},
  that such a measure grows exponentially with
  lattice size, $\eta(\lambda^*) \propto exp(\alpha_* L)$, with $\alpha_* \sim 2.36\times10^{-3}$. 
     Notice however that, despite such exponentially increasing behavior, at the  thermodynamic
 limit the period $\tau$ diverges, and there is no revival of the function, consequently, the non-Markovianity pointed by this measure must be null:
  $\eta(\lambda^*) = 0$ for $L\rightarrow \infty $. It should be clear by now, that the non-Markovianity we have observed so far is due to the finite size
  of the lattice and  the  periodical dynamical revivals thereof.
The behavior of the Loschmidt echo outside of the critical point 
 is plotted in Fig. \ref{fig.losch.ncp}. We highlight some of  its features: 
(i) due to finite size effects, we 
see that after a certain time ($\Gamma$), which increases 
with the lattice size ($\Gamma \propto L$), the 
function has a chaotic behavior; (ii) the ``shape'' of 
the function before the chaotic behavior is invariant with 
the lattice size, only its amplitude is changed.

\begin{figure}[H]
\includegraphics[width=9.4cm]{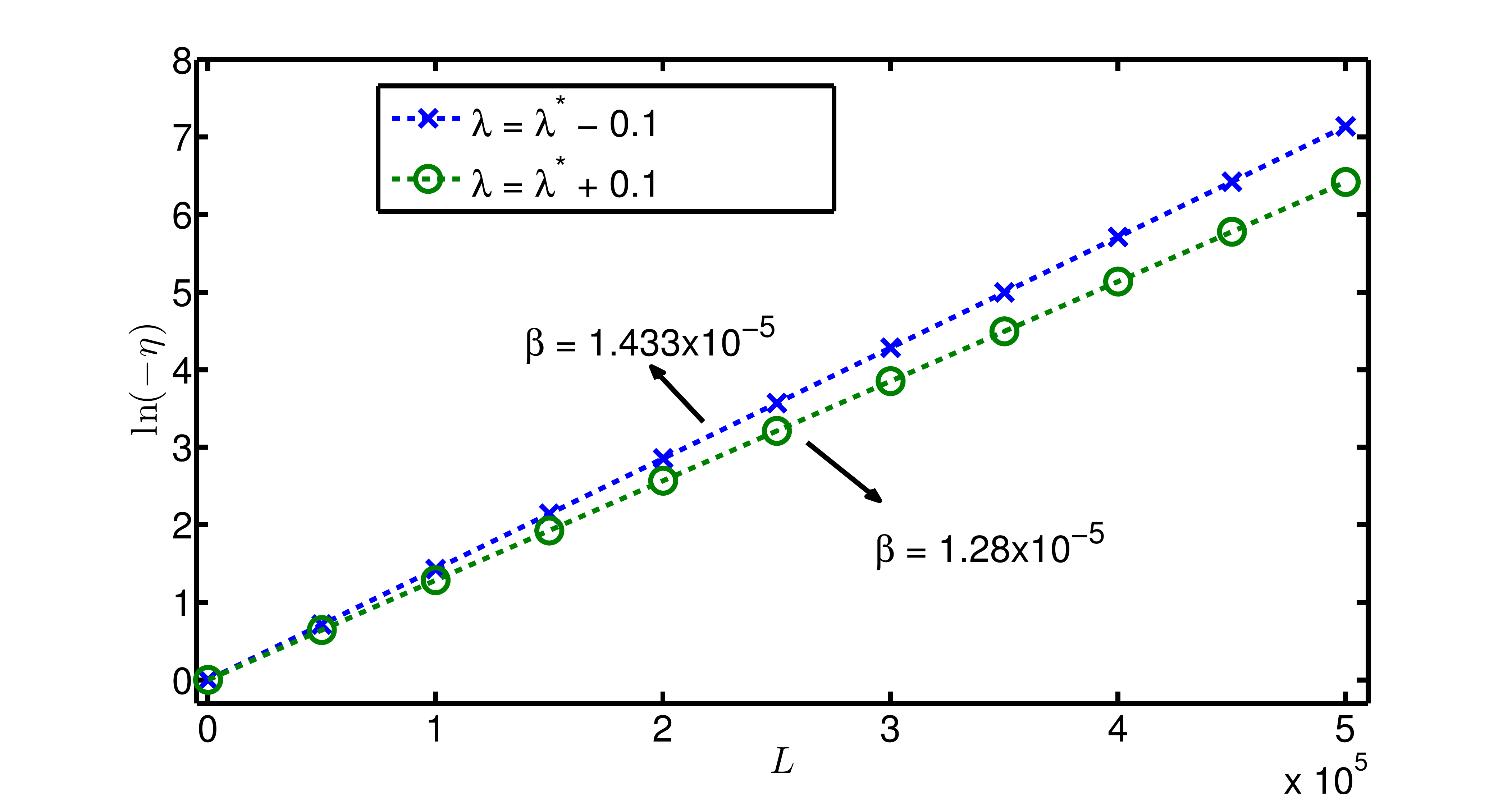}
 \caption{Finite size scaling analysis: $ln(-\eta)$ as a function 
of $L$, for $L=100$ to $L=5\times10^5$ sites, outside of  the critical point, more precisely,
 for $\lambda = \lambda^* \pm 0.1 $, and $\delta=10^{-2}$.
The linear fit reveals an exponential
divergence of the non-Markovianity with the lattice size, $(-\eta) \propto e^{\beta L}$.
\label{fig.Nscaling.ncp} }
\end{figure}

Performing 
then a finite-size scaling analysis, we see, in Fig. \ref{fig.Nscaling.ncp}, that the
 non-Markovianity measure grows exponentially with lattice size, 
 $\eta(\lambda^* - 0.1) \propto exp(\beta_l L)$, with $\beta_l \sim 1.43\times10^{-5}$, 
 and $\eta(\lambda^* + 0.1) \propto exp(\beta_r L)$, with $\beta_r \sim 1.29\times10^{-5}$.
  Notice that although  the measure also has an exponential scaling, as in the critical point,
   its exponential factors are much smaller, namely, $ \beta_{l(r)} / {\alpha_*} \sim 10^{-2}$.

In summary,  we see that the non-Markovianity measure, for finite size systems, 
reaches its maximal at the critical point, whereas in the thermodynamic 
limit it is zero exactly at the critical point, and it  diverges outside of  the critical point.

Assuming the environment described by  the Ising Hamiltonian, the measure $(\eta)$ (Eq.\ref{measureN}) and the witness $(\mathcal{N})$ (Eq.\ref{nmq})
for the non-Markovian dynamics for the two qubits  have
exactly  the
 same behavior of the non-Markovian dynamics for one qubit. Here we will just highlight that the results do not depend on the parameters $J_{S}$ and $\lambda_{S}$, and 
  the choice  of a Hamiltonian $H_S$ for the open system (two spins) just adds a  relative phase in its initial state, $\ket{\chi(0)}$, do not affecting $(\eta)$ nor $(\mathcal{N})$.

\section{Entanglement as a witness of non-Markovianity in the Ising model: Beyond finite size effects}

In the previous section, we characterized the non-Markovianity  by means of the non-positivity of the dynamical matrix expressed as a simple function of the Loschmidt echo. Now we will further explore the dynamics using a witness of non-Markovianity. Different non-Markovianity witnesses based 
on entanglement, or on bipartite correlations, have recently appeared in the 
literature \cite{Rivas_2010,Luo_2012,debarba}. We based our witness on the entanglement between the central  
qubit coupled to an ancilla.
 Our main concern shall  be to detect
the non-Markovianity that is not due to the finite lattice size.
To see how this works, we assume a system $S$, with dynamics described by a map $\Phi$, and a static ancillary system $A$. The system-ancilla evolution is given by, 
\begin{equation}
\rho_{SA}(t_f)= \Phi(t_f,t_0)\otimes\mathbb{I}_A\left[\rho_{SA}(t_0)\right].
\end{equation} 
Note that we have trivially extended the map to a separable one, with  no local action over the ancilla.

Entanglement cannot be generated by a local CP map. Assuming that the map ($\Phi(t_f,t_0))$ is divisible, in the sense discussed in section II, i.e., the intermediate map $(\Phi(t_f,t_m))$ is CP, $t_f > t_m > t_0$, we have:
\begin{eqnarray}\label{ent}
E\left[\rho_{SA}(t_f)\right] &=& E\left[(\Phi(t_f,t_m)\Phi(t_m,t_0)\otimes\mathbb{I}_A\left[\rho_{SA}(t_0)\right]\right]\nonumber\\
&=& E\left[(\Phi(t_f,t_m)\otimes\mathbb{I}_A\left[\rho_{SA}(t_m)\right]\right]\nonumber\\
&\le&E\left[\rho_{SA}(t_m)\right],
\end{eqnarray} 
where $E\left[\rho_{SA}(t)\right]$ is some quantifier of bipartite entanglement. The above equation expresses the fact that entanglement is
monotonically decreasing under local CP maps. 

In order to simplify notation, from now on  we shall write $E\left[\rho_{SA}(t)\right]=E_{SA}(t)$. From Eq.(\ref{ent}) we have that a local CP divisible map leads to a monotonic decrease $\left(\frac{d}{dt}E_{SA}(t)\le 0\right)$ of an entanglement measure of the system and  ancilla. Therefore any violation of this monotonicity $\left(\frac{d}{dt}E_{SA}(t)> 0\right)$ is a  sufficient criterion to witness non-Markovianity. Based on this idea, we can consider a witness  $(\mathcal{N})$ of non-Markovianity  in the form \cite{Rivas_2010}:
\begin{equation}\label{nmq}
\mathcal{N}=\int_{(d/dt) E_{SA}>0}\frac{d}{dt}E_{SA}(t), 
\end{equation} 
such  that  $\mathcal{N}>0$ for  non-Markovian dynamics.

 Now consider  system  and ancilla as two qubits in an initial  maximally entangled state, $\ket{\phi^{+}}=\left(\ket{gg}+\ket{ee}\right)/\sqrt{2}$. The system is under the action of
  the map given by Eq.(\ref{phit0}), and the ancilla is let alone. We resume the  study of our  problem (Fig. 1) under this new perspective.

  \begin{figure}[H]
\includegraphics[width=9.0cm]{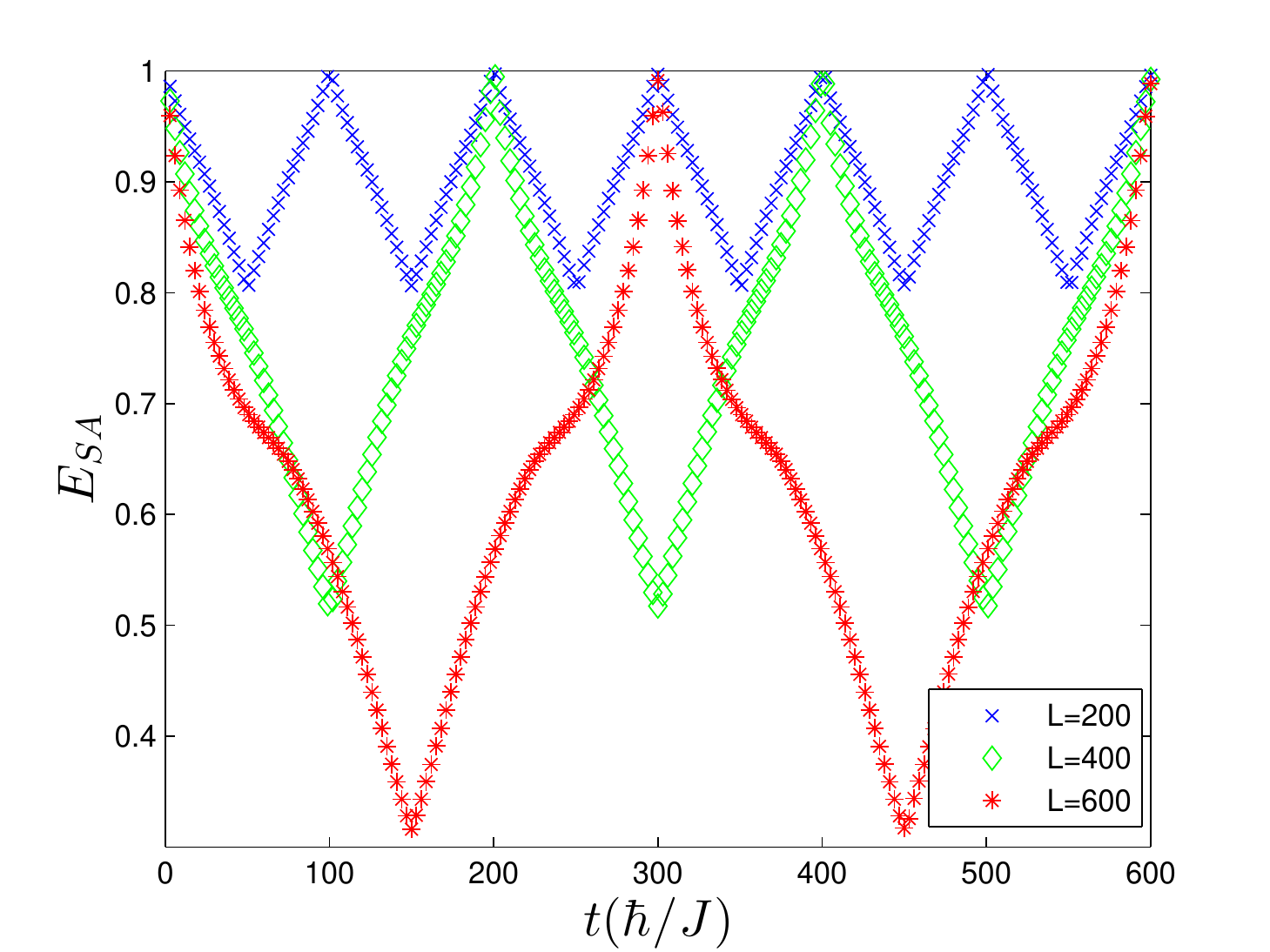}
  \caption{ The entanglement measure $E_{SA}$, quantified  by the negativity as a function of time, at the critical point $\lambda^*= \lambda + \delta = 1$, and $\delta=10^{-2}$,
  for different lattice sizes.
 }
\end{figure}

In Fig.4,  we saw that at  the critical point $\lambda^{*}=1$, the revival of the Loschmidt echo, {\it i.e.} the revival of the coherence (recoherence), occurs in a time $\tau$ proportional to the lattice size.  
This  non-Markovianity, due to  the finite size of the lattice,  allows for  the open system to regain coherence and information from the environment. It is  shown in Fig. 8, where  the entanglement measure ($E_{SA}$ ) is the negativity,  for different lattice sizes at the critical point. The period of time in which  the negativity increases is proportional to the lattice size,
 as expected. 
However, looking  at outside of the critical point,  in a time before the detection of non-Markovianity due to  the size effect, we can witness non-Markovianity related to the characteristic features of the environment. This fact was  observed before by means of the distinguishability of two quantum states \cite{Haikka12}. In Fig. 9,  we plot the negativity, for different lattice sizes, outside of the critical point,  with fixed interaction coupling  constant $\delta=0.01$, in  a time interval  excluding the finite size effect. We see that  even for different 
lattice  sizes  the negativity presents the same behavior, {\it i.e.} the period of time in which  $E_{SA}$ monotonically  increases is the same. The degree of non-Markovianity, quantified by $\mathcal{N}$ (Eq.(\ref{nmq})), becomes higher as we increase the lattice size, $\mathcal{N}=\sum_n\left(E_{SA}(\tau_{n}^{max})-E_{SA}(\tau_{n}^{min})\right)$, where $E_{SA}(\tau_{n}^{max})$ and $E_{SA}(\tau_{n}^{min})$ are the set of local maximum and minimum values of $E_{SA}(t)$. At this point one  can  note that the behavior of the negativity is similar  to the Loschmidt echo, more precisely, in this specific case we have the interesting result:
\begin{equation}
E_{SA}=\sqrt{\mathcal{L}}.
\end{equation}
The above equation follows from the definition of negativity,  $E_{SA}=\sum_i\left(|p_i|-p_i\right)$, where the $p_i$  are the four  eigenvalues $\frac{1}{2}(-|x(t)|,|x(t)|,1,1)$ of
 $\rho^{\Gamma}_{SA}(t)$,  which is the partial trace of $\rho_{SA}(t)=\Phi(t,0)\otimes\mathbb{I}_A\left[\ketbra{\phi^{+}}{\phi^{+}}\right]=(\ketbra{g}{g} \otimes \ketbra{g}{g} + \ketbra{e}{e} \otimes \ketbra{e}{e} +\ketbra{g}{e} \otimes \ketbra{g}{e} x(t)^* + \ketbra{e}{g} \otimes \ketbra{e}{g} x(t))/2$. 
In Fig.10, we see the witness of non-Markovianity, against the effective transverse field $(\lambda_{ef}=\lambda+\delta)$, for two different lattice sizes, in an interval that 
avoids finite size effects. Increasing the field from small values, the witness decreases, until it gets close to the critical point,  where it starts to increase, and  suddenly 
drops to zero,  exactly  at the critical point ($\mathcal{N}(\lambda^{*})=0$). This is a very nice result to conclude this section, for the dynamics is known to be Markovian at the critical point.

\begin{figure}[H]
\includegraphics[width=9.0cm]{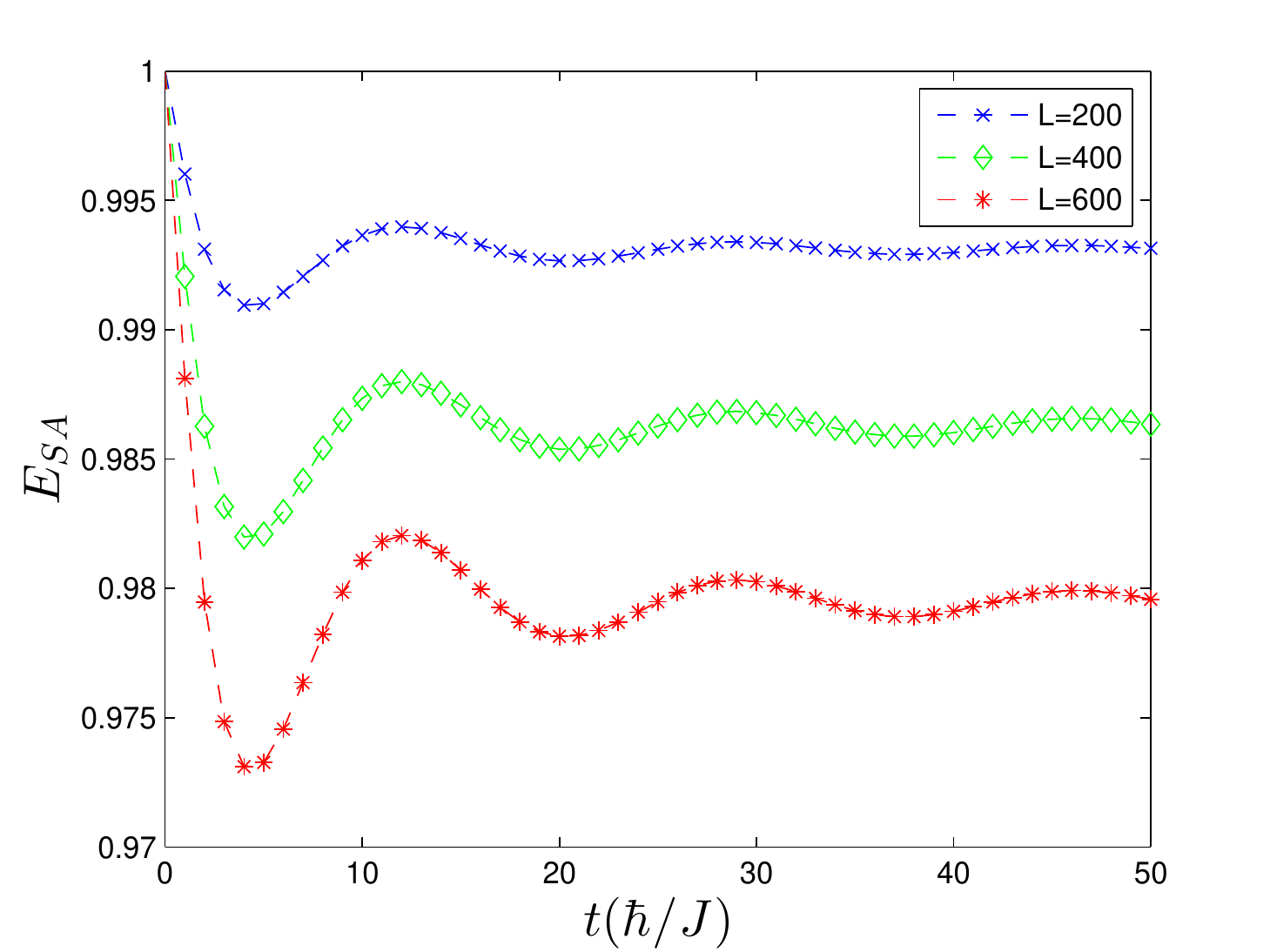}
  \caption{The negativity $E_{SA}$ as a function 
of  time, outside of the critical point, for $\lambda = \lambda^* - 0.1 $, and $\delta=10^{-2}$, for different lattice sizes.
 }
\end{figure}

\begin{figure}[H]
\includegraphics[width=9.0cm]{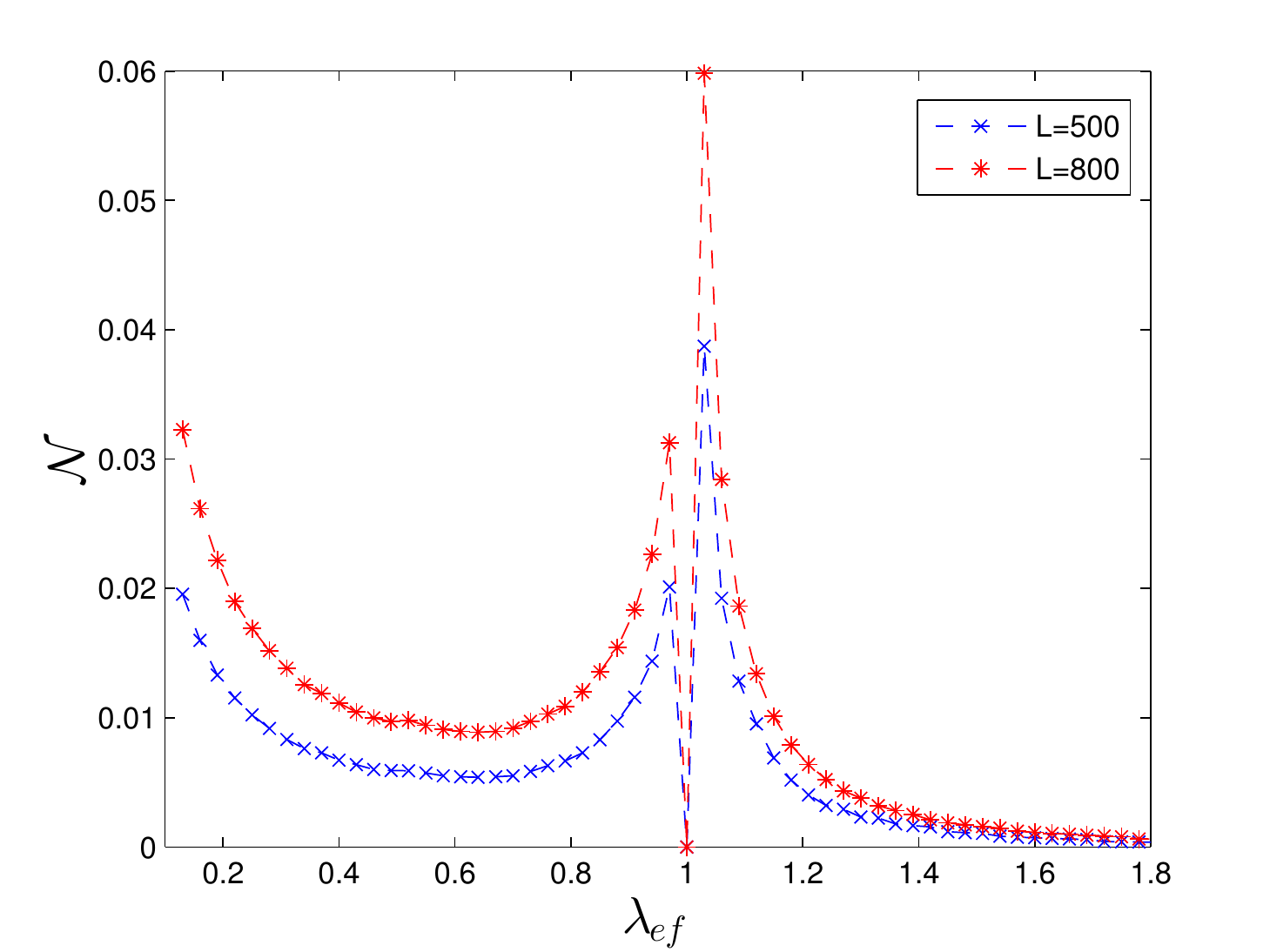}
  \caption{ The witness of non-Markovianity   $\mathcal{N}$ as a function of the
effective field $\lambda_{ef}=\lambda+\delta$, for $\delta=0.01$, and in a time window  excluding finite size effects.
 }
\end{figure}
\section{Conclusion}
\label{conclusion}
We derived the  analytical expression for the
Kraus representation of  the map corresponding to the evolution of one and two  qubits interacting
with an environment represented by a general   quadratic fermionic Hamiltonian. We concluded that the non-Markovian dynamics of two qubits interacting with the Ising environment
does not present any new feature in relation to  the dynamics of one qubit.
We introduced simple
functions to check the non-Markovianity 
of the dynamics.
For the particular case of the Ising environment, 
 we investigated the dynamics of one qubit interacting with  lattices up to $10^5$ sites.
  We quantified the non-Markovianity
by the most negative eigenvalue ($\eta$ - Eq.\ref{measureN}) of the dynamical 
matrix, and obtained that,
 for finite size systems, 
it reaches its maximum at the critical point, whereas in the thermodynamic 
limit it is zero exactly at the critical point, diverging outside of the critical point. 
We also quantified the non-Markovianity using an entanglement based approach ($\mathcal{N}$ - Eq.\ref{nmq} ). We showed, in the case of  one qubit interacting with Ising model, that the non-Markovianity measures we introduced are simple functions of
the Loschmidt echo.
Finally, we clearly identified two kinds of 
non-Markovianity, one due to the finite size of the environment, and another intrinsic of the Ising Hamiltonian, and
we were able to quantify both.

\acknowledgments 
We acknowledge financial support by the Brazilian agencies INCT-IQ (National Institute of Science and
Technology for Quantum Information), FAPEMIG, and CNPq. 

\subsection*{ Author contribution statement } Fernando Iemini and Leonardo da Silva Souza are the
main authors and equally contributed with development
of algorithms, numerical calculations, analytical calculations, data analysis 
and the art of figures. T. Debarba, A.T. Cesário, T. Maciel and R.O. Vianna helped with the
calculations, data analysis, interpretation of the results.

\end{document}